\documentclass[12pt]{article}
\usepackage{amsfonts,amsmath,amssymb}

\def\hybrid{\topmargin 0pt      \oddsidemargin 0pt
    \headheight 0pt \headsep 0pt

       \textwidth 6.5in        
       \textheight 9in         
       \marginparwidth 0.0in
       \parskip 5pt plus 1pt   \jot = 1.5ex}

\catcode`\@=11
\def\numberbysection{\@addtoreset{equation}{section}
        \def\theequation{\thesection.\arabic{equation}}}

\def\underline#1{\relax\ifmmode\@@underline#1\else
        $\@@underline{\hbox{#1}}$\relax\fi}

\def\titlepage{\@restonecolfalse\if@twocolumn\@restonecoltrue\onecolumn
     \else \newpage \fi \thispagestyle{empty}\c@page\z@
        \def\thefootnote{\fnsymbol{footnote}} }

\def\endtitlepage{\if@restonecol\twocolumn \else  \fi
        \def\thefootnote{\arabic{footnote}}
        \setcounter{footnote}{0}}  

\relax
\numberbysection
\hybrid
\newtheorem{teo}{Theorem}[section]

\newtheorem{cor}{Corollary}[section]
\newtheorem{lem}{Lemma}[section]

\newenvironment{dok}{\noindent{\it Proof.}}{\hfill$\square$}
\newcommand{\ie}{i.\,e.}
\newcommand{\DE}{Darboux\,--\,Egoroff}
\newcommand{\BA}{Baker\,--\,Akhiezer}

\newcommand{\handB}{{\cal B}}
\newcommand{\handR}{{\cal R}}
\newcommand{\handD}{{\cal D}}
\newcommand{\handC}{{\cal C}}
\newcommand{\handF}{{\cal F}}

\newcommand{\handK}{{\cal K}}

\newcommand{\bst}{\beta^{*}}
\newcommand{\va}{{\bf a}}
\newcommand{\vb}{{\bf b}}
\newcommand{\vc}{{\bf c}}
\newcommand{\vd}{{\bf d}}
\newcommand{\vl}{{\bf l}}
\newcommand{\vr}{{\bf r}}
\newcommand{\vx}{{\bf x}}

\renewcommand{\phi}{\varphi}
\newcommand{\eps}{\varepsilon}
\newcommand{\ds}{\displaystyle}
\newcommand{\wt}{\widetilde}
\newcommand{\wh}{\widehat}
\newcommand{\ov}{\overline}
\newcommand{\inte}{\mathbb Z}
\newcommand{\real}{\mathbb R}
\newcommand{\comp}{\mathbb C}
\newcommand{\sca}[1]{\langle #1\rangle}

\newcommand{\Res}{\mathop{\rm Res}\nolimits}
\newcommand{\sref}[2]{(\ref{#1},\,\ref{#2})}
\newcommand{\lref}[2]{(\ref{#1}${}-{}$\ref{#2})}
\newcommand{\wT}{\widetilde T}
\newcommand{\p}{\partial}
\newcommand{\D}{\Delta}

\begin{document}

\begin{titlepage}

\title{Discrete analogs of the \DE\ metrics}

\author{A.\,A.\,Akhmetshin{}${}^*$
\and
I.\,M.\,Krichever{}${}^*$
\and
Y.\,S.\,Volvovski\thanks{Research supported in part by the National 
Science Foundation under grant DMS-98-02577}}
\date{December 15, 1998}
\maketitle

\begin{center}
${}^*$Columbia University, 2990 Broadway, New York, NY 10027, USA\\
and\\
Landau Institute for Theoretical Physics,\\
2 Kosygina str., Moscow, 117940, Russia.\\
\end{center}
\vspace{0.5in}

\begin{abstract}
Discrete analogs of the \DE\ metrics are considered.
It is shown that the corresponding lattices in the Euclidean
space are described by a set of functions $h_i^{\pm}(u)$, $u\in\inte^n$.
Discrete analogs of the Lam\'e equations are found. It is proved that 
up to a gauge transformation these equations are necessary and sufficient 
for discrete analogs of rotation coefficients to exist.
Explicit examples of the \DE\ lattices are constructed by means of 
algebro-geometric methods.
\end{abstract}

\end{titlepage}
\newpage

\section{Introduction}

Discrete analogs of various special coordinate systems on two-dimensional 
surfaces in three-dimensional Euclidean space, and discrete analogs of 
multi-dimensional conjugated coordinate nets have attracted great interest 
recently \cite{bob,dol1,dol2}.

This interest has been motivated by revealed connections between 
corresponding problems of classical (continuous) differential geometry 
and modern problems of mathematical and theoretical physics. 
For example, it turned out that one of the central problems of 
the differential geometry of the previous century: the problem of 
constructing $n$-orthogonal curvilinear coordinate systems, 
or \emph{flat} diagonal metrics
\begin{equation}\label{1}
ds^2=\sum_{i=1}^n H_i^2(u) (du^i)^2,\qquad u=(u^1,\ldots,u^n),
\end{equation}
is deeply connected to the theory of integrable quasilinear $(1+1)$-systems 
of the hydrodynamic type. These systems are the core of the~Whitham 
approach to the perturbation theory of soliton equations \cite{dn1,dn2,tsar}. 
Moreover, as it was noticed in \cite{dub1}, the classification problem of 
the so-called \DE\ metrics, \ie\ flat diagonal metrics such that
\begin{equation}\label{2}
\p_j H_i^2=\p_i H_j^2,\qquad \p_i=\frac{\p}{\p u^i},
\end{equation}
is equivalent to the classification problem of massive topological quantum
field models~\cite{wdvv,wdvv1,wdvv2}.

It should be emphasized that classical results \cite{dar} in the theory of 
$n$-orthogonal curvilinear coordinate systems were mainly of classification 
nature, and as a result the list of explicit examples of such coordinate 
systems was relatively short. In the remarkable paper~\cite{zak} it was 
shown that a wide class of solutions to the Lam\'e equations which describe 
the rotation coefficients
\begin{equation}\label{5}
\beta_{ij}=\frac{\p_iH_j}{H_i},\qquad i\neq j,
\end{equation}
of flat diagonal metrics, can be obtained with the help of the ``dressing'' 
procedure which is well-known in the theory of solitons. 
Results of~\cite{zak,zak2} were a starting point of the work~\cite{kr} 
by one of the authors, where a construction of algebro-geometric 
$n$-orthogonal coordinate systems was proposed, and a new type of solutions 
of the associativity equations was found. Explicit expressions in terms of 
the Riemann theta-functions associated with auxiliary algebraic curves 
were obtained.

Originally the main goal of this paper was to construct \emph{discrete}
analogs of the algebro-geometric $n$-orthogonal coordinate systems.
Note, that in general the problem of finding an \emph{integrable} discrete 
analog of integrable continuous system is ill-defined and has no universal 
answer. At the same time discretization methods developed in the theory of 
solitions are universal enough to be applicable to all the systems that are 
considered in the framework of the inverse method. They use natural 
discretization of auxiliary linear problems or even more natural change of 
analytical properties with respect to spectral parameter of common 
eigenfunctions of auxiliary linear problems.

It turns out that discretization of the algebro-geometric scheme of~\cite{kr} 
leads to a construction of lattices of vectors $\vx(u)=(x^1(u),\dots,x^n(u))$ 
in the Euclidean space parameterized by integer $n$-dimensional vectors
$u=(u^1,\dots,u^n)$, $u^i\in\inte$, and satisfying \emph{planar} and 
\emph{circular} conditions. These conditions were proposed in~\cite{dol3} 
as discrete analogs of general $n$-orthogonal coordinate systems. 
Note, that planar and circular lattices in three-dimensional space were 
introduced for the first time in \cite{bob2}, based on the earlier 
works \cite{sau} and \cite{cs1,cs2}.

The planar condition means that for each pair of indices $i$, $j$ the 
corresponding elementary quadrilateral of the lattice, \ie\ the polygon with 
vertices $\{\vx(u),\,T_i\vx(u),\,T_j\vx(u),\,T_iT_j\vx(u)\}$, is flat. 
Here and below $T_i$ denotes the shift operator in the discrete 
variable $u^i$: 
$$
T_i\vx(u^1,\dots,u^i,\dots,u^n)=\vx(u^1,\dots,u^i+1,\dots,u^n).
$$
The circular condition means that each of the elementary quadrilatteral
can be inscribed into a circle, \ie\ that the sum of opposite angles of
the polygon equals~$\pi$.

The main goal of the paper is to prove integrability of the lattices 
satisfying more rigid constraint than the circular condition.
Namely, that for each pair of indices $i\neq j$ the edges of the lattice
\begin{equation}\label{edg}
X_i^+(u)=T_i\vx(u)-\vx(u),\quad
X_j^-(u)=T_j^{-1}\vx(u)-\vx(u)
\end{equation}
with vertices $\{\vx(u),\,T_i\vx(u)\}$ and $\{\vx(u),\,T_j^{-1}\vx(u)\}$
are orthogonal to each other, \ie\
\begin{equation}\label{or}
\sca{X_i^+,X_j^-}=0.
\end{equation}
Here and below $\sca{{}\cdot{},\!{}\cdot{}}$ stands for Euclidean scalar 
product of $n$-dimensional vectors. Note that (\ref{or}) implies that 
the two opposite angles of the polygon are right and therefore, 
the corresponding lattice satisfies the circular condition.

Lattices satisfying (\ref{or}) will be called \DE\ lattices, because 
(\ref{or}) implies in particular, that there exists a potential function 
$\Phi(u)$ such that its discrete derivatives equal to the lengths of edges:
\begin{equation}\label{phi}
\D_i \Phi(u)=\sca{X_i^+(u), X_i^+(u)}.
\end{equation}
Note that in the continuous case the definition (\ref{2}) of the \DE\ metric
is equivalent to the existence of function~$\Phi$ such that
$\p_i\Phi=H_i^2=\sca{\p_ix,\p_ix}$.

Constraint (\ref{or}) has naturally arisen from the discretization of
the algebro-geometric scheme~\cite{kr}%
\footnote{When the reduction (\ref{or}) was found the authors became aware
that the same reduction was proposed in \cite{dol4}.}.
Moreover, the algebro-geometric construction of \DE\ lattices suggests 
the possibility to introduce discrete analogs of the Lam\'e coefficients
$h_i^{\pm}(u)$. It should be mentioned that unlike the continuous case 
the definition of such coefficients is local, but not ultra-local and 
requires scalar products of edges not from the same but also from the nearest 
vertices. As a result, it turns out that the proof that these coefficients 
are well-defined is not evident.

This proof is presented in the next section of the paper. In the third 
section we derive a full set of equations which describe the discrete 
Lam\'e coefficients and prove their integrability. Algebro-geometric 
construction of the \DE\ lattices is presented in the last section.

We use the following notation for the discrete derivatives:
$$
\D_iF(u)=T_iF(u)-F(u),\qquad \D^-_iF(u)=T^-_iF(u)-F(u),
\qquad T^-_i=T^{-1}_i .
$$
Various objects in the paper has upper indices~$+$ and~$-$. 
Sometimes for the sake of brevity we omit the index~$+$, 
\ie\ we assume that $F(u)=F^+(u)$. 
We use also the following discrete analogs of the Leibnitz rule
$$
\begin{array}{rcl}
\D_i(F(u)G(u))&=&\D_iF(u)\,G(u)+T_iF(u)\,\D_iG(u)={}\\[3mm]
{}&=&\D_iF(u)\,G(u)+F(u)\,\D_iG(u)+\D_iF(u)\,\D_iG(u)
\end{array}
$$
and the formula for taking discrete derivatives of a ratio
$$
\D_i\left(\frac{F(u)}{G(u)}\right)=
\frac{\D_iF(u)\,G(u)-F(u)\,\D_iG(u)}{G(u)\,T_iG(u)}.
$$
At the end of the introduction we would like to emphasize that our 
definition of the discrete Lam\'e coefficients extensively uses the 
properties of the \DE\ lattices. The problem of a similar description 
of \emph{intrinsic} geometry of general analogs of flat diagonal metrics 
is still open. We would like to consider this problem and more general 
problem of intrinsic geometry on graphs in future.

\section{Discrete analogs of the Lam\'e coefficients}

To begin with let us give an equivalent definition of the \DE\ lattices. 
This definition uses only the orthogonality properties of edges of the 
lattice. Note, that all the lattices in this paper are assumed to be 
non-degenerate, \ie\ for any $u$ the vectors $X_i(u)$, $i=1,\dots,n$, 
are linearly independent.

\begin{lem}\label{ldef}
The following definitions $1^0-3^0$ of the \DE\ lattice $\vx(u)$
are equivalent:

\noindent
$1^0.$ The lattice $\vx(u)$ is planar and for any $i\ne j$ equation
(\ref{or}) holds:
$$
\sca{X_i,X_j^{-}}=0.
$$

\noindent
$2^0.$ For any triple of indices $i$, $j$, $m$ different from each other
we have
\begin{equation}\label{or2}
\sca{X_i^+, X_j^-}=0,\qquad  \sca{T_m X_i^+, X_j^-}=0.
\end{equation}

\noindent
$3^0.$ For any $j\ne i$ and any set $\{m_1,m_2,\dots,m_s\}$ of distinct
indices which does not contain~$i$ it follows
\begin{equation}\label{or3}
\sca{X_i^+, X_j^-}=0,\qquad
\sca{\Bigl(\textstyle\prod\limits_{k=1}^s T_{m_k}\Bigr) X_i^+, X_j^-}=0 .
\end{equation}
\end{lem}
\begin{dok}
Since $\vx(u)$ is planar we see that the vector $T_m X_i$ is a linear
combination of the vectors $X_i$ and $X_m$. From (\ref{or}) it follows
that $T_m X_i$ is orthogonal to $X_j^-$. Therefore, the equality 
(\ref{or2}) follows from the definition $1^0$.

In the reverse case, from (\ref{or2}) it follows that the vectors $X_i$,
$X_j$ and~$T_i X_j$ are orthogonal to all the vectors $X_m^-$, $m\ne i,j$.
Since the lattice is non-degenerate due to our assumption we obtain
that $X_i$, $X_j$ and~$T_i X_j$ belong to the two-dimensional plane,
which is orthogonal to all vectors $X_m^-$, $m\ne i,j$.
This implies that the lattice is planar.

We have proved that definitions $1^0$ and $2^0$ are equivalent. In the
same way one can prove the equivalence of definitions $2^0$ and $3^0$.
\end{dok}

\begin{teo}\label{tmin}
For any \DE\ lattice $\vx(u)$ there exists a unique set of functions
$h_i^{\pm}(u)$, $i=1,\dots,n$, normalized by the condition 
$h^{+}_i(0,\dots,0)=1$, and such that the following equalities for 
the scalar products of edges of the lattice hold
\begin{eqnarray}
\sca{X_i,X_i}&=&2(T_i h_i^{+})\cdot h_i^{-};              \label{m1}\\
\sca{X_i,X^{-}_i}&=&-(T_i h_i^{+})\cdot(T_i^{-} h_i^{-}); \label{m2}\\
\sca{T_j X_i,X^{-}_i}&=&-(T_i T_j h_i^{+})\cdot(T^{-}_i h_i^{-}),
\quad i\neq j.                                            \label{m3}
\end{eqnarray}
\end{teo}
Note that equalities \lref{m1}{m3} are invariant under the gauge 
transformation $h_i^{\pm}(u)\mapsto a_i^{\pm 1}h_i^{\pm}$, where $\{a_i\}$ 
is a set of arbitrary nonzero constants.
Initial conditions $h_i^{+}(0)=1$ are chosen to fix the gauge.

\begin{dok}
Let us fix some index $i$. First we shall show that equations 
\lref{m1}{m3} imply Pfaff-type system of partial difference equations 
for the function $h_i^+$. 
Using \sref{m1}{m2} we obtain two different expressions for $h_i^-$:
\begin{equation}\label{i1}
h_i^{-}=\frac{\sca{X_i,X_i}}{2T_i h_i^{+}}=
\frac{\sca{T_i X_i,X_i}}{T_i T_i h_i^{+}}.
\end{equation}
(Here and later we use equality $T_iX_i^{-}=-X_i$, which follows directly 
from the definition of~$X_i^{\pm}$).
It follows from (\ref{i1}) that $h_i^+$ satisfies the difference equation
\begin{equation}\label{i2}
T_i h_i^{+}=-2h_i^{+} \frac{\sca{X_i,X_i^-}}{\sca{X_i^-,X_i^-}}
           =-2h_i^{+}\cdot A .
\end{equation}
If we replace $h_i^-$ in (\ref{m3}) by the middle term of (\ref{i1})
we obtain
\begin{equation}\label{ij}
T_jT_i h_i^{+}=-2h_i^{+} \frac{\sca{T_j X_i,X_i^-}}{\sca{X_i^-,X_i^-}}
=-2h_i^{+}\cdot B_j.
\end{equation}
Let us show that equations (\ref{i2}) and (\ref{ij}) are compatible.
The compatibility conditions are equivalent to the equation
$\,T_j A\cdot T_i^{-1}B_j = B_j\cdot T_i^{-1}A$, that is
\begin{equation}\label{ij2}
T_j \left( \frac{\sca{X_i,X_i^-}}{\sca{X_i^-,X_i^-}} \right)\cdot
T_i^{-1}\left( \frac{\sca{T_jX_i,X_i^-}}{\sca{X_i^-,X_i^-}} \right)
=\frac{\sca{T_jX_i,X_i^-}}{\sca{X_i^-,X_i^-}} \cdot
T_i^{-1} \left( \frac{\sca{X_i,X_i^-}}{\sca{X_i^-,X_i^-}} \right) .
\end{equation}
Applying $T_i$ to the both sides we get
\begin{equation}\label{ij3}
\frac{\sca{T_jT_iX_i,T_jX_i}\cdot\sca{T_jX_i,X_i^-}}{|T_jX_i|^2}
=\frac{\sca{T_jT_iX_i,X_i}\cdot\sca{X_i,X_i^-}}{|X_i|^2} .
\end{equation}
Let us denote the vectors $X_i^-$, $X_i$, $T_jX_i$, $T_iT_jX_i$ by
$\va$, $\vb$, $\vc$ and $\vd$ respectively (see fig.\,\ref{pic1}).
Consider orthogonal projections $\va'$, $\vd'$ of the vectors 
$\va$, $\vd$ on the plane spanned by the edges~$X_i$ and $X_j$ 
(the plane of the central plaquet in fig.\,\ref{pic1}).
Replacing $\va$, $\vd$ by $\va'$, $\vd'$ in equation (\ref{ij3}),
we change no scalar product in it. Dividing both sides by $|\va'| |\vd'|$, 
we obtain
$$
\frac{\sca{\va',\vc}\,\sca{\vc,\vd'}}{|\va'| |\vc|^2 |\vd'|}=
\cos(\va',\vc) \cos(\vc,\vd')=
\cos(\va',\vb) \cos(\vb,\vd')=
\frac{\sca{\va',\vb}\,\sca{\vb,\vd'}}{|\va'| |\vb|^2 |\vd'|} .
$$
It follows from the definition $1^0$ of the \DE\ lattice that the vectors 
$\va'$ and $\vc$ are both orthogonal to the vector $\vl=X_j$.
Similarly, $\vb$ and~$\vd'$ are both orthogonal to $\vr=T_iX_j$.
Consequently,
$$
\cos(\va',\vc)=\cos(\vb,\vd')=1,\qquad \cos(\va',\vb)= \cos(\vc,\vd').
$$
The latter equalities prove compatiblity of (\ref{i2}) and (\ref{ij}).

\begin{figure}
\unitlength=2mm
\begin{center}
\begin{picture}(40,18)
 \put(2,8){\circle*{0.5}}
 \put(0,16){\circle*{0.5}}
 \put(8,8){\circle*{0.5}}
 \put(8,18){\circle*{0.5}}
 \put(20,6){\circle*{0.5}}
 \put(22,18){\circle*{0.5}}
 \put(43,14.5){\circle*{0.5}}
 \put(35,-1.5){\circle*{0.5}}
 \put(2,8){\vector(1,0){6}}
 \put(2,8){\line(-1,4){2}}
 \put(0,16){\line(4,1){8}}
 \put(8,8){\line(0,1){10}}
 \put(8,8){\line(6,-1){12}}
 \put(8,18){\vector(1,0){14}}
 \put(20,6){\line(1,6){2}}
 \put(20,6){\line(2,-1){15}}
 \put(22,18){\line(6,-1){21}}
 \put(35,-1.5){\line(1,2){8}}
 \put(19.6,6){\vector(4,-1){0.1}}
 \put(42.6,14.5){\vector(4,-1){0.1}}
 \put(4.5,6.2){$\va$}
 \put(13.2,5){$\vb$}
 \put(14.5,18.5){$\vc$}
 \put(32,17){$\vd$}
 \put(8.5,8.5){{$\vx(u)$}}
 \put(2,18.5){{$T_j\vx(u)$}}
 \put(20.5,6.2){{$T_i\vx(u)$}}
 \put(6.7,12){$\vl$}
 \put(21.5,12){$\vr$}
\end{picture}
\end{center}
\caption{}\label{pic1}
\end{figure}

Let us consider now two equations (\ref {ij}) for indices $j$ and $k$
such that $i\ne j\ne k$. In terms of the coefficients $B_j$ and $B_k$
of these equations their compatibility is equivalent to the equality
$B_k\cdot T_k B_j = B_j\cdot T_j B_k$, which can be written as follows
\begin{equation}\label{ijk}
\frac{\sca{T_k X_i^{-},T_i^{-}X_i^{-}}\cdot
      \sca{T_k T_j X_i,T_k X_i^-}}{|T_k X_i^{-}|^2}=
\frac{\sca{T_j X_i^{-},T_i^{-}X_i^-}\cdot
      \sca{T_j T_k X_i,T_j X_i^-}}{|T_j X_i^{-}|^2}\ .
\end{equation}
Let us introduce the following notations: 
$\va=T_i^{-}X_i^{-}$, $\vb=T_k X_i^{-}$, $\vc=T_j X_i^{-}$, 
$\vd=T_k T_j X_i$ (see fig.\,\ref{pic2}). 
Let $\handC$ be the three-dimensional cube with edges $T_i^{-}X_i$,
$T_i^{-}X_j$ and $T_i^{-}X_k$ (the central cube in fig.\,\ref{pic2}).
Let $\va'$ and $\vd'$ be orthogonal projections of the vectors $\va$ 
and~$\vd$ on the three-dimensional space spanned by the edges 
of $\handC$. One can replace $\va$, $\vd$ by $\va'$, $\vd'$, 
respectively, in (\ref{ijk}) without changing scalar products in it.
Moreover, equality (\ref{ijk}) remains valid if the vectors $\va'$ 
and $\vd'$ are multiplied by any non-zero constants.

From the orthogonality properties of the \DE\ lattices it follows that
$\va'$ is proportional to $\wt\va=T_j T_k X_i^{-}$, because both of them 
are orthogonal to the plane  $\pi_L$, spanned by $T_i^{-}X_j$ and 
$T_i^{-}X_k$ (left face of $\handC$). In the same way, the vector $\vd'$ 
is proportional to~$\wt\vd=X_i^{-}$, because they are orthogonal to the
plane~$\pi_R$, spanned by $X_j$ and $X_k$ (right face of~$\handC$). 
Hence, in (\ref{ijk}) we may replace $\va$ and $\vd$ by the vectors 
$\wt\va$ and $\wt\vd$. If we divide it by $|\wt\va| |\wt\vd|$, 
it takes the form
$$
\frac{\sca{\wt\va,\vc}\,\sca{\vc,\wt\vd}}{|\wt\va| |\vc|^2 |\wt\vd|}
=\cos(\wt\va,\vc) \cos(\vc,\wt\vd)=
\cos(\wt\va,\vb) \cos(\vb,\wt\vd)=
\frac{\sca{\wt\va,\vb}\,\sca{\vb,\wt\vd}}{|\wt\va| |\vb|^2 |\wt\vd|}\  .
$$
To prove the last equality we use the following result, which is
well-known in spherical geometry:
\begin{lem}
Let one of the dihedral angles of a three-edged piramid be a right angle. 
Let $\alpha_1$, $\alpha_2$ and~$\alpha_3$ be the plane angles at the vertex, 
and $\alpha_3$ corresponds to the face opposite to the right dihedral angle. 
Then
$$
\cos\alpha_3=\cos\alpha_1 \cos\alpha_2 .
$$
\end{lem}

The upper face of the central cube $\handC$ (the face with edges 
$T_jT_i^- X_i$ and $T_jT_i^- X_k$) is perpendicular to the front face
(the face with edges $T_i^- X_i$ and $T_i^- X_j$). At the same time,
the lower face of $\handC$ (the face with edges $T_i^- X_i$ and $T_i^- X_k$)
is perpendicular to the rear face (the face with edges $T_kT_i^- X_i$ and
$T_k T_i^- X_j$). Therefore, the above presented lemma implies
$$ 
\cos(\wt\va,\vb) \cos(\vb,\wt\vd)=
\cos(\wt\va,\wt\vd)= \cos(\wt\va,\vc) \cos(\vc,\wt\vd) .
$$
Hence, the relation (\ref{ijk}), and, consequently,
compatibility of (\ref{m3}) for distinct $i$ and $j$ are proved.

\begin{figure}
\unitlength=3mm
\begin{center}
\begin{picture}(32,15)
 \put(2,2){\circle*{0.5}}
 \put(8,2){\circle*{0.5}}
 \put(0,10){\circle*{0.5}}
 \put(8,12){\circle*{0.5}}
 \put(5,5){\circle*{0.5}}
 \put(11,5){\circle*{0.5}}
 \put(3,13){\circle*{0.5}}
 \put(11,15){\circle*{0.5}}
 \put(18,2){\circle*{0.5}}
 \put(21,5){\circle*{0.5}}
 \put(18,12){\circle*{0.5}}
 \put(21,15){\circle*{0.5}}
 \put(30,0){\circle*{0.5}}
 \put(33,3){\circle*{0.5}}
 \put(32,12){\circle*{0.5}}
 \put(35,15){\circle*{0.5}}
 \put(8,2){\vector(-1,0){6}}
 \put(18,2){\vector(-1,0){10}}
 \put(18,2){\line(6,-1){12}}
 \put(0,10){\line(4,1){8}}
 \put(18,12){\vector(-1,0){10}}
 \put(18,12){\line(1,0){14}}
 \put(3,13){\line(4,1){8}}
 \put(21,15){\vector(-1,0){10}}
 \put(35,15){\vector(-1,0){14}}
 \put(2,2){\line(-1,4){2}}
 \put(8,2){\line(0,1){10}}
 \put(18,2){\line(0,1){10}}
 \put(30,0){\line(1,6){2}}
 \put(33,3){\line(1,6){2}}
 \put(0,10){\line(1,1){3}}
 \put(8,12){\line(1,1){3}}
 \put(18,12){\line(1,1){3}}
 \put(32,12){\line(1,1){3}}
 \put(30,0){\line(1,1){3}}
 \put(2,2){\line(1,1){3}}
 \put(8,2){\line(1,1){3}}
 \put(18,2){\line(1,1){3}}
 \put(5,5){\line(1,0){6}}
 \put(5,5){\line(-1,4){2}}
 \put(11,5){\line(0,1){10}}
 \put(21,5){\line(0,1){10}}
 \put(21,5){\line(6,-1){12}}
 \put(21,5){\vector(-1,0){10}}
 \put(4.5,1){{\small $\va$}}
 \put(15,5.3){{\small $\vb$}}
 \put(12.5,12.3){{\small $\vc$}}
 \put(27,15.3){{\small $\vd$}}
 \put(15.5,15.3){{\small $\wt\va$}}
 \put(12,0.5){{\small $\wt\vd$}}
 \put(15.5,0.8){{\small $\vx(u)$}}
 \put(30.2,-1.2){{\small $T_i\vx(u)$}}
 \put(21.3,5.3){{\small $T_k\vx(u)$}}
 \put(14.1,10.7){{\small $T_j\vx(u)$}}
 \put(9,8){{\small $\pi_L$}}
 \put(19,8){{\small $\pi_R$}}
\end{picture}
\end{center}
\caption{}\label{pic2}
\end{figure}

Equations (\ref{i2}) and (\ref{ij}) uniquely define $h_i^+(u)$, if we fix 
the normalization $h_i^{+}(0)=1$. Note, that the functions $h_i^{\pm}(u)$ 
with given~$i$ are defined independently of $h_j^{\pm}(u)$ with other 
indices. Theorem is proved.
\end{dok}

The definition of $h_i^{\pm}$ uses a minimal set of scalar products. 
It turns out that this set is \emph{complete} in the following sense.
All the other scalar products of the edges can be expressed through the
scalar products from the minimal set, and therefore, through the
functions $h_i^{\pm}$, $i=1,\dots,n$. Let us present some of the
corresponding formulae which will be used later.

\begin{lem}\label{le1}
Let $h_i^{\pm}(u)$, $i=1,\dots,n$, be the functions
defined by Theorem \ref{tmin}. Then
\begin{align}
\sca{X_i,X_j}&=-2\frac{(\D_iT_jh_j^+)\, (T_ih_i^+)\, h_j^-}{T_iT_jh_i^+},
\qquad i\ne j,                                 \label{sp0}\\
\sca{X_i^-,X_j^-}&=-2\frac{(\D_i^- T_j^- h_j^-)\, (T_i^- h_i^-)\, h_j^+}%
{T_i^- T_j^- h_i^-},\qquad i\ne j.             \label{sm0}
\end{align}
\end{lem}
\begin{dok}
Due to planar property of the lattice, $T_jX_i$ is a linear combination 
of~$X_i$ and $X_j$, \ie\
\begin{equation}\label{pl}
T_jX_i=\alpha X_i+\beta X_j .
\end{equation}
Let us find the coefficients of the sum. Taking scalar product
of (\ref{pl}) with~$X_i^-$, we get a relation which implies
$$
\alpha=\frac{\sca{T_jX_i,X_i^-}}{\sca{X_i,X_i^-}}
      =\frac{T_iT_jh_i^+}{T_ih_i^+}.
$$
If we take scalar product of (\ref{pl}) with $X_j^-$ and use the equality
$T_jX_i=T_iX_j + X_i-X_j$, then we get the formula
$$
\beta=\frac{\sca{T_iX_j,X_j^-}-\sca{X_j,X_j^-}}{\sca{X_j,X_j^-}}
     =\frac{\D_iT_jh_j^+}{T_jh_j^+}.
$$
Let us multiply (\ref{pl}) by the vector $X_j$, which is orthogonal 
to~$T_jX_i$. We get $\alpha\sca{X_i,X_j}+\beta\sca{X_j,X_j}=0$, 
which proves (\ref{sp0}):
$$
\sca{X_i,X_j}=
\left( \frac{\sca{T_iX_j,X_j^-}}{\sca{X_j,X_j^-}}-1 \right)\cdot
\frac{\sca{X_i,X_i^-}}{\sca{T_jX_i,X_i^-}}\cdot |X_j^2|
=-2\frac{(\D_iT_jh_j^+)\, (T_ih_i^+)\, h_j^-}{T_iT_jh_i^+} .
$$
The second formula (\ref{sm0}) can be proved in the same way.
\end{dok}

It turns out that the scalar products $\sca{X_i^{\pm},X_j^{\pm}}$ can 
be expressed through the Lam\'e coefficients by different formulae.
\begin{lem}\label{le2}
For any \DE\ lattice the following formulae
\begin{align}
\sca{X_i,X_j}&=-2(T_i h_i^+) (\D_j h_i^-),          \label{sp}\\
\sca{X_i^-,X_j^-}&=-2(T_i^- h_i^-) (\D_j^- h_i^+),  \label{sm}
\end{align} 
are valid. Here $h_i^{\pm}(u)$, $i=1,\dots,n$, are the functions 
defined by Theorem \ref{tmin}.
\end{lem}
\begin{dok}
Let us express $\sca{X_i,X_j}$ through the scalar products from
the minimal set \lref{m1}{m3}. 
From the definition of \DE\ lattices it follows that
$0=\sca{X_i,T_iX_j}=\sca{X_i,X_j}-\sca{X_i,X_i}+\sca{X_i,T_jX_i}.$
In order to express the last term in the right hand side of this equality
through the minimal set, note that~$T_jX_i$ is proportional to the 
orthogonal projection~$\vd$ of the vector~$X_i^{-}$ on the plane spanned 
by $X_i$ and~$X_j$. Therefore,
$$
\sca{X_i,T_jX_i}=\sca{X_i,X_i^-}\frac{|T_jX_i|}{|\vd|}=
\sca{X_i,X_i^-}\frac{\sca{T_jX_i,T_jX_i}}{\sca{T_jX_i,X_i^-}}\ .
$$
We obtain
$$
\sca{X_i,X_j}=|X_i|^2-\sca{X_i,X_i^-}\frac{|T_jX_i|^2}{\sca{T_jX_i,X_i^-}}.
$$
Direct substitution of \lref{m1}{m3} into the latter formula
gives~(\ref{sp}). Formula~(\ref{sm}) can be proved in the same way.
\end{dok}

Lemmas \ref{le1} and \ref{le2} imply constraints for the functions 
$h_i^{\pm}(u)$. Using the symmetry of scalar product we get from 
(\ref{sp}) and (\ref{sm}) the equations:
\begin{equation}\label{bpm}
\frac{\D_ih_j^-}{T_ih_i^+}=\frac{\D_jh_i^-}{T_jh_j^+},\qquad
\frac{\D_i^- h_j^+}{T_i^- h_i^-}=\frac{\D_j^- h_i^+}{T_j^- h_j^-},
\qquad i\ne j.
\end{equation}
From (\ref{sp0}) and (\ref{sp}) it follows that
\begin{equation}\label{b1}
\beta_{ij}^+=-\beta_{ji}^- , 
\end{equation}
where
\begin{equation}\label{b}
\beta^{+}_{ij}=\frac{\D_ih_j}{T_ih_i},\qquad
\beta^{-}_{ji}= \frac{\D_j^- h_i^-}{T_j^-h_j^-},\qquad i\ne j.
\end{equation}
The functions $\beta^{\pm}_{ij}(u)$ are discrete analogs of the rotation 
coefficients (\ref{5}) of flat diagonal metrics. Below we will often omit 
upper index~$+$ for the functions $\beta^{+}_{ij}(u)$.

Relations (\ref{bpm}) and (\ref{b1}) are discrete analogs of symmetry
conditions for \DE\ metrics. For $n\geq 3$, there is another analog of 
symmetry conditions which can be written in terms of the functions 
$\beta_{ij}$ only.
\begin{lem}
Let $\beta_{ij}(u)$ be the discrete rotation coefficients of \DE\ lattice. 
Then for each triple of pairwise distinct indices $i$, $j$, $k$ the equation
\begin{equation}\label{bbb}
(T_k\beta_{ik})\, (T_i\beta_{ji})\, (T_j\beta_{kj})=
(T_k\beta_{jk})\, (T_j\beta_{ij})\, (T_i\beta_{ki}).
\end{equation}
is fulfilled.
\end{lem}
\begin{dok}
From (\ref{sp0}) and definition of the discrete rotation coefficients it
follows that
$$
\sca{X_i,X_k}=(T_k\beta_{ik})\cdot\frac{h_k^-}{T_k h_k^+}\,.
$$
The desired relation can be obtained if we divide the equality
$\sca{X_i,X_k}\sca{X_j,X_i}\sca{X_k,X_j}=
\sca{X_j,X_k}\sca{X_i,X_j}\sca{X_k,X_i}$
by symmetric expression
$$
\frac{h_i^-}{T_i h_i^+}\cdot\frac{h_j^-}{T_jh_j^+}\cdot
\frac{h_k^-}{T_k h_k^+}\,.
$$
\end{dok}

Of course, constraints (\ref{bpm}) and (\ref{b1}) do not form a complete 
set of equations on the discrete Lam\'e coefficients. The main goal 
of the next section is to get such a set of equations.

\section{Discrete analogs of the Lam\'e equations}

Just as in the continuous case the discrete analog of the Lam\'e equations
can be obtained as compatibility conditions of a system of linear equations
satisfied by the edges $X_i^{\pm}(u)$ of the lattice. These linear 
equations become simplier after an appropriate rescaling of the edges.

For each \DE\ lattice let us define vectors $Y_i^{\pm}(u)$ by the formulae
\begin{equation}\label{y}
Y_i(u)=\frac{1}{T_ih_i^{+}(u)} X_i(u),\qquad
Y_i^-(u)=\frac{1}{T_i^{-}h_i^{-}(u)} X_i^{-}(u),
\end{equation}
where $h_i^{\pm}(u)$ are the functions constructed in Theorem \ref{tmin}.
Equation (\ref{m2}) implies that these vectors form biorthogonal system:
\begin{equation}\label{y1}
\sca{Y_i,Y_j^-}=-\delta_{ij}. 
\end{equation}
For further needs we write down some other scalar products
which can be obtained from the relations \lref{m1}{m3}:
\begin{equation}\label{y2}
\sca{Y_i,Y_i}=2\frac{h_i^-}{T_ih_i^+},\qquad
\sca{T_iY_i,Y_i}=\frac{h_i^-}{T_ih_i^+},\qquad \sca{T_jY_i,Y_i^-}=-1,
\end{equation}
and \sref{sp0}{sp}:
\begin{equation}\label{y3}
\sca{Y_i,Y_j}=-2\frac{(\D_iT_jh_j^+)\, h_j^-}{(T_iT_jh_i^+)\, (T_jh_j^+)}
=-2\frac{\Delta_jh_i^-}{T_jh_j^+} \ .
\end{equation}
(The analogous formula can be written for $\sca{Y_i^-,Y_j^-}$.)

\begin{teo}
For any \DE\ lattice the vectors $Y^{\pm}_i(u)$ satisfy the system of
equations:
\begin{align}
\D^{\pm}_j Y^{\pm}_i&=(T^{\pm}_j\beta^{\pm}_{ij}) Y^{\pm}_j,
\qquad i\neq j,                                \label{l1}\\
\D^{\pm}_i Y^{\pm}_i&=b^{\pm}_iY^{\pm}_i-
\sum_{j\ne i}\beta^{\pm}_{ij}Y^{\pm}_j,        \label{l2}
\end{align}
where the coefficients $\beta^{\pm}_{ij}(u)$ are defined by (\ref{b}),
and the coefficients $b_i^{\pm}(u)$ are given by formulae
\begin{equation}\label{v}
b_i^{+}=-\frac{1}{2}-\sum_{j\ne i}\beta_{ij}(T_i\beta_{ji}),\qquad
b_i^{-}=-\frac{1}{2}-\sum_{j\ne i}\beta^{-}_{ij}(T_i^-\beta^{-}_{ji}) \ .
\end{equation}
\end{teo}
\begin{dok}
Since the lattice is planar, the vector $\Delta_jY_i$ ($i\ne j$) is a linear 
combination of $Y_i$ and $Y_j$, \ie\ $\Delta_jY_i=a_{ij}Y_i+d_{ij}Y_j$.
To find the coefficient~$a_{ij}$, we take scalar product of this equation 
with the vector~$Y_i^-$. Using (\ref{y1}) and (\ref{y2}), we obtain
$a_{ij}=\sca{T_jY_i-Y_i, Y_i^-}=0.$ Therefore, $\Delta_jY_i=d_{ij}Y_j$.
Taking the scalar product of this equation and the vector~$Y_j$ we obtain
$$
\sca{\Delta_jY_i,Y_j}=d_{ij}\sca{Y_j,Y_j}=2d_{ij}\,
\frac{h_j^-}{T_jh_j^+}\ .
$$
On the other hand, (\ref{y3}) implies
$$
\sca{\Delta_jY_i,Y_j}=\sca{T_jY_i,Y_j}-\sca{Y_i,Y_j}=-\sca{Y_i,Y_j}=
2T_j\left(\frac{\D_ih_j^+}{T_ih_i^+}\right)\frac{h_j^-}{T_jh_j^+}\,.
$$
Comparison of the right hand sides of the two last equations yields
$d_{ij}=T_j \beta_{ij}$. In the same way we can prove (\ref{l1}) for 
the vectors~$Y_i^{-}$.

We will show now that $Y_i(u)$ satisfy (\ref{l2}). For any $u$ the 
vectors $Y_1(u),\dots,Y_n(u)$ form a basis of~$\real^n$ (since the 
lattice is non-degenerate). Thus, there exists a decomposition
\begin{equation}\label{l3}
\D_iY_i=b_i Y_i+\sum_{j\ne i} c_{ij} Y_j.
\end{equation}
Coefficient $c_{ij}$ can be found by taking the scalar product of this 
equation with the vector~$Y_j^-$. We get 
$\sca{\D_iY_i,Y_j^-}=c_{ij}\sca{Y_jY_j^-}=-c_{ij}.$
On the other hand
$$
\sca{\D_iY_i,Y_j^-}=\D_i\sca{Y_i,Y_j^-}-\sca{T_iY_i,\D_iY_j^-}=
T_i\sca{Y_i,\D_i^-Y_j^-}.
$$
Using (\ref{l1}), which has been just proved, we obtain
$$
-c_{ij}=\sca{\D_iY_i,Y_j^-}=T_i\sca{Y_i,(T_i^-\beta^{-}_{ji})Y_i^-}=
-\beta^{-}_{ji}=\beta_{ij}.
$$
To determine $b_i(u),$ we multiply (\ref{l3}) by $Y_i$:
\begin{equation}\label{l4}
\sca{\D_iY_i,Y_i}=b_i\sca{Y_i,Y_i}-\sum_{i\ne j}\beta_{ij}\sca{Y_i,Y_j}.
\end{equation}
Plugging (\ref{y2}) and (\ref{y3}) into equation (\ref{l4}) we establish
formula (\ref{v}) for $b_i(u)$. The second equation in (\ref{l2}) can be
obtained in the same manner.
\end{dok}

The compatibility conditions of the linear system \sref{l1}{l2} are 
discrete analogs of the Lam\'e equations on the rotation coefficients of 
the \DE\ metrics. Our next goal is to prove the inverse statement: any 
solution of these equations uniquely defines the rotation coefficients 
of some \DE\ lattice.

\begin{teo}\label{tbs}
Let functions $\beta_{ij}(u)$, $i\ne j$, $i,j\in\{1,\dots,n\}$ 
\emph{($n\geq 3$)}, satisfy the relations (\ref{bbb})
$$
(T_k\beta_{ik})\, (T_i\beta_{ji})\, (T_j\beta_{kj})=
(T_k\beta_{jk})\, (T_j\beta_{ij})\, (T_i\beta_{ki})
$$
and the equations
\begin{align}
\D_k\beta_{ij}&=(T_k\beta_{ik})\beta_{kj},\qquad i\ne j\ne k,  \label{so1}\\
\D_jv_i&=\D_i(\beta_{ji}\, T_j\beta_{ij}),\qquad i\ne j,       \label{so2}\\
\D_i\D_j\beta_{ij}+\D_i\beta_{ij}+\D_j\beta_{ij}&=T_j\beta_{ij}(T_jv_i-v_j)-
\sum\limits_{k\ne i,j} T_j(\beta_{ik}\beta_{kj}),\quad i\ne j,  \label{so3}
\end{align}
where $v_i(u)$ is defined by
\begin{equation}\label{so4}
v_i=-\sum_{p\ne i}\beta_{ip}(T_i\beta_{pi})\, .
\end{equation}
Then there exists a unique constant $\eta$ such that the functions
\begin{equation}\label{so5}
\wt\beta_{ij}(u)=(2\eta)^{u_j-u_i-1}\beta_{ij}(u)
\end{equation}
are the rotation coefficients of some \DE\ lattice.
\end{teo}

Equations \lref{so1}{so3} are equivalent to compatibility conditions of the
linear system
\begin{align}
\D_j \Psi_i&=(T_j\beta_{ij}) \Psi_j,\qquad i\neq j,          \label{ll1}\\
\D_i \Psi_i&=(\mu+v_i)\Psi_i-\sum_{j\ne i}\beta_{ij}\Psi_j\, \label{ll2}
\end{align}
where $\mu$ is an arbitrary complex constant. The transformation (\ref{so5})
sends solutions of \lref{so1}{so4} to the solutions of the same system and
preserves the relations (\ref{bbb}), since it corresponds to the gauge 
transformation
\begin{equation}
Y_i=\wt\Psi_i=(2\eta)^{-u_i}\Psi_i \label{kal}
\end{equation}
of the linear system \sref{ll1}{ll2}.
Therefore, the theorem gives the necessary and sufficient conditions for 
the functions $\beta_{ij}(u)$ to be the rotation coefficients of some 
\DE\ lattice, up to gauge transformation.
\medskip

\begin{dok}
Unlike the continuous case, the problem of reconstruction the \DE\ lattice
from the functions $\beta_{ij}$  requires a highly nontrivial choice of the 
initial data for the solutions of system \sref{l1}{l2}. We fix these data
by defining the matrix of scalar products of the corresponding vectors.
In the next two lemmas we shall construct functions $\bst_{ij}(u)$
by  means of which we shall later determine these scalar products.

\begin{lem}\label{lbii}
Let functions $\beta_{ij}(u)$ satisfy the conditions (\ref{bbb}) and
\lref{so1}{so4}. Then there exists a unique solution $f_i(u)$, $i=1,\dots,n$,
of the system
\begin{align}
\D_k f_i&=-(T_k\beta_{ik})\, (T_i\beta_{ki}) f_i, \qquad i\ne k,
                                                         \label{sf1}\\
(T_i\beta_{ki}) f_i&=(T_k\beta_{ik}) f_k, \qquad i\ne k, \label{sf2}
\end{align}
normalized by the condition $f_1(0)=1$.
\end{lem}

\begin{dok}
System (\ref{sf1}, \ref{sf2}) is over-determined, but we will show that it
is equivalent to a system of $n$ compatible equations on the
function $f_1(u)$. Note, that equations
(\ref{sf2}) are compatible due to (\ref{bbb}).

First of all, let us prove the compatibility of a pair of equations 
(\ref{sf1}) with distinct values of the index $k$, say $p$ and $q$. 
It suffices to show that the following expression for $\D_p \D_q f_i$ 
is symmetric with respect to $p$ and $q$:
$$
\D_p\D_q f_i=-\Bigl[ \D_p(T_q\beta_{iq}\cdot T_i\beta_{qi})-
T_p(T_q\beta_{iq}\cdot T_i\beta_{qi})\,
(T_p\beta_{ip}\cdot T_i\beta_{pi}) \Bigr] f_i .
$$
Indeed, due to (\ref{so1}) the expression in brackets is equal to
$$
\begin{array}{c}
[\,\ldots\,]=T_q(T_p\beta_{ip}\cdot\beta_{pq}) (T_i\beta_{qi})+
(T_pT_q\beta_{iq})\, T_i(T_p\beta_{qp}\cdot\beta_{pi})-
(T_pT_q\beta_{iq})\, T_p(T_i\beta_{qi}\cdot\beta_{ip})\,
  (T_i\beta_{pi})={}\\[3mm]
{}=(T_qT_p\beta_{ip})(T_q\beta_{pq})(T_i\beta_{qi})+(T_pT_q\beta_{iq})\,   
[T_iT_p\beta_{qp}-T_p\D_i\beta_{qp}]\, (T_i\beta_{pi}) .
\end{array}
$$
Obviously, the last formula has the symmetry needed.

Next, let us consider two equations (\ref{sf1}) with distinct values of the
index~$i$ (denote them again by~$p$ and~$q$). We shall show that one of them 
implies the other provided~$f_p$ and~$f_q$ satisfy (\ref{sf2}). Indeed,
$$
\begin{array}{l}
\D_kf_p=\D_k \left( \ds\frac{T_q\beta_{pq}}{T_p\beta_{qp}} f_q \right)=
\D_k \left( \ds\frac{T_q\beta_{pq}}{T_p\beta_{qp}} \right) f_q -
\ds T_k\left(\frac{T_q\beta_{pq}}{T_p\beta_{qp}}\right) (T_k\beta_{qk})\,
(T_q\beta_{kq})\,f_q={}\\[5mm]
\quad{}=\ds\left( T_q(T_k\beta_{pk}\cdot\beta_{kq}) -
\frac{T_q\beta_{pq}}{T_p\beta_{qp}}T_p(T_k\beta_{qk}\cdot\beta_{kp})-
T_k(T_q\beta_{pq}\cdot\beta_{qk})\,(T_q\beta_{kq}) \right)\,
\frac{f_q}{T_kT_p\beta_{qp}}={}\\[5mm]
\quad{}=\ds\left( (T_q\beta_{kq})(T_k\beta_{pk})-
\frac{T_q\beta_{pq}}{T_p\beta_{qp}} (T_pT_k\beta_{qk})(T_p\beta_{kp})
\right) \frac{f_q}{T_kT_p\beta_{qp}} .
\end{array}
$$
Now, (\ref{bbb}) implies
$(T_q\beta_{kq})(T_k\beta_{pk})f_q=(T_k\beta_{qk})(T_p\beta_{kp})f_p$.
Finally, we have
$$
\D_kf_p=\Bigl( (T_k\beta_{qk})(T_p\beta_{kp})-
(T_pT_k\beta_{qk})(T_p\beta_{kp}) \Bigr) \frac{f_p}{T_kT_p\beta_{qp}}
=-(T_k\beta_{pk})\,(T_p\beta_{kp})\,f_p ,
$$
which is the formula we need.

Now let us consider equation (\ref{sf1}) for $\D_1 f_i$, $i>1$.
Plugging into both sides of this equation the expression
$f_i=\alpha_{i1}f_1$, with $\alpha_{i1}=(T_1\beta_{i1})/(T_i\beta_{1i})$,
we obtain the equation of the type $\D_1 f_1=\handF_i(u)f_1$, where the
function $\handF_i(u)$ is a rational combination of $\beta_{1i}(u)$ and
$\beta_{i1}(u)$. In fact, the function $\handF_i(u)$ and, therefore, 
the equation on $\D_1 f_1$ do not depend on the index~$i$. 
In order to prove that, we choose index $j\ne i$ and consider equation 
(\ref{sf1}) for~$\D_1 f_j$. It was shown at the second step of this 
proof that the substitution $f_j=\alpha_{ji}f_i$ leads to (\ref{sf1}) 
for $\D_1 f_i$. On the other hand, substitution $f_j=\alpha_{j1}f_1$ 
gives the equation $\D_1 f_1=\handF_j(u)f_1$. 
Therefore, the equation $\D_1 f_1=\handF_i(u)f_1$ can be obtained 
from $\D_1 f_1=\handF_j(u)f_1$ by substitution 
$f_1=(\alpha_{j1})^{-1} \alpha_{ji}\alpha_{i1} f_1$. 
But due to (\ref{bbb}), 
$(\alpha_{j1})^{-1}\alpha_{ji}\alpha_{i1}=
\alpha_{1j}\alpha_{ji}\alpha_{i1}=1$, 
and we conclude that $\handF_i(u)=\handF_j(u)=\handF(u)$.

Thus we obtain $n$ equations on the function $f_1(u)$, namely, the equations
\begin{equation}\label{eqf}
\D_1f_1=\handF(u) f_1,\qquad
\D_if_1=-(T_1\beta_{i1}) (T_i\beta_{1i}) f_1,\quad i=2,\dots,n.
\end{equation}
It was already shown that the equations from the second group are compatible.
To establish compatibility of the equations $\D_1f_1=\handF(u) f_1$
and $\D_if_1=-(T_1\beta_{i1}) (T_i\beta_{1i}) f_1$, $i>1$,
it suffices to note that they are gauge equivalent to compatible equations
$\D_1f_j=-(T_1\beta_{j1}) (T_j\beta_{1j}) f_j$ and
$\D_if_j=-(T_j\beta_{ij}) (T_i\beta_{ji}) f_j$ for any index $j$ not
equal to $1$ and $i$.

Summarizing all the facts, we see that the solution of system \lref{sf1}{sf2} 
can be obtained as follows. First, we define the function $f_1(u)$ from 
compatible system (\ref{eqf}) with the initial condition $f_1(0)=1$. 
Then, using (\ref{sf2}), we obtain all the other functions $f_i(u)$, 
$i=2,\dots,n$. Then, as it was shown at the second step of the proof, 
all the equations of the system are fulfilled. 
The proof of lemma is now complete.
\end{dok}

Let us define $\bst_{ii}(u)$ by $\bst_{ii}(u)=f_i(u)$, $i=1,\dots,n$,
where $f_i(u)$ were constructed in the previous lemma.
For $i\ne j$ we define $\bst_{ij}$ by the formula
\begin{equation}\label{ss2}
-\bst_{ij}=(T_j\beta_{ij})\,\bst_{jj},\qquad j\ne i.
\end{equation}
Note, that (\ref{sf2}) implies $\bst_{ij}(u)=\bst_{ji}(u)$.

\begin{lem}
The above-defined functions $\bst_{ij}(u)$, $i,j\in\{1,\dots,n\}$,
satisfy the following system of equations:
\begin{align}
\D_j\bst_{ik}&=(T_j\beta_{ij})\,\bst_{jk},\qquad j\ne i,k,  \label{ss1}\\
\D_i\bst_{ij}&=(v_i-1)\bst_{ij}-\sum_{k\ne i}\beta_{ik}\bst_{kj},
\qquad i\ne j .                                             \label{ss3}
\end{align}
\end{lem}

\begin{dok}
Lemma is proved by straightforward calculations. Further in this proof we
assume that $i\neq j\neq k$. Note, that lemma \ref{lbii} and the definition 
of $\bst_{ii}$ imply:
$$
\D_j\bst_{ii}=-(T_j\beta_{ij})(T_i\beta_{ji})\bst_{ii}=
(T_j\beta_{ij})\bst_{ji}.
$$
Now, by definition
$\D_j\bst_{ki}=-\D_j(T_i\beta_{ki}\cdot\bst_{ii})=-(T_jT_i\beta_{ki})
\D_j\bst_{ii}-\D_j(T_i\beta_{ki})\bst_{ii}.$
Using the above established equalities and (\ref{so1}), we get
$$
\begin{aligned}
\D_j\bst_{ki}&=(T_jT_i\beta_{ki})(T_j\beta_{ij})(T_i\beta_{ji})\bst_{ii}-
(T_jT_i\beta_{kj})(T_i\beta_{ji})\bst_{ii}={}\\[3mm]
{}&=(\D_iT_j\beta_{kj})(T_i\beta_{ji})\bst_{ii}-
(T_jT_i\beta_{kj})(T_i\beta_{ji})\bst_{ii}=
-(T_j\beta_{kj})(T_i\beta_{ji})\bst_{ii}=(T_j\beta_{kj})\bst_{ji}.
\end{aligned}
$$
Thus, we establish equations (\ref{ss1}).
Let us prove (\ref{ss3}). Using (\ref{so3}), we obtain
$$
\begin{array}{rcl}
\D_j\bst_{ji}&=&-\D_j(T_i\beta_{ji}\cdot\bst_{ii})=
  -(T_jT_i\beta_{ji})\D_j\bst_{ii}-(\D_jT_i\beta_{ji})\bst_{ii}={}\\[3mm]
{}&=&\ds (T_jT_i\beta_{ji})(T_j\beta_{ij})(T_i\beta_{ji})\bst_{ii}-
  \Bigl[ T_i\beta_{ji}(-v_i+T_iv_j)-\sum\limits_{k\neq i,j}T_i(\beta_{jk}
  \beta_{ki})-\D_i\beta_{ji} \Bigr] \bst_{ii}={}\\[5mm]
{}&=&\ds \Bigl( T_i\beta_{ji}[v_i-T_iv_j+(T_jT_i\beta_{ji})T_j\beta_{ij}]
  +\sum\limits_{k\neq i,j}T_i(\beta_{jk}\beta_{ki})+\D_i\beta_{ji} \Bigr)
  \bst_{ii}={}\\[5mm]
{}&=&\ds -\bst_{ji}(v_i-T_iv_j+T_j(T_i\beta_{ji}\cdot\beta_{ij}))
-\sum\limits_{k\neq i,j}(T_i\beta_{jk})\bst_{ki}-\bst_{ji}-\beta_{ji}\bst_{ii}.
\end{array}
$$
Plugging into the last formula $\D+1$ instead of the shift operators~$T$ 
and applying (\ref{so2}) we have
$$
\begin{array}{l}
\D_j\bst_{ji}=\ds -\bst_{ji}\Bigl( v_i-v_j+(T_i\beta_{ji})\beta_{ij}\Bigr) -
  \sum\limits_{k\neq j}\beta_{jk}\bst_{ki}
  -\sum\limits_{k\neq i,j}(\D_i\beta_{jk})\bst_{ki}-\bst_{ji}={}\\[5mm]
\quad{}=\ds -\bst_{ji}(v_i+1-v_j+(T_i\beta_{ji})\beta_{ij})
  -\sum\limits_{k\neq j}\beta_{jk}\bst_{ki}
  -\sum\limits_{k\neq i,j}(T_i\beta_{ji}\cdot\beta_{ik})\bst_{ki}={}\\[5mm]
\quad{}=\ds \sum\limits_{k\neq i,j}(T_i\beta_{ki}\cdot\beta_{ik})\bst_{ji}
  -\sum\limits_{k\neq i,j}(T_i\beta_{ji}\cdot\beta_{ik})\bst_{ki}
  -\sum\limits_{k\neq j}\beta_{jk}\bst_{ki}+(v_j-1)\bst_{ji}={}\\[5mm]
\quad{}=\ds -\sum\limits_{k\neq i,j}\left[
  (T_i\beta_{ki}\cdot\beta_{ik})(T_i\beta_{ji})\bst_{ii}
  -(T_i\beta_{ji}\cdot\beta_{ik})(T_i\beta_{ki})\bst_{ii}\right]
  -\sum\limits_{k\neq j} \beta_{jk}\bst_{ki}+(v_j-1)\bst_{ji}={}\\[5mm]
\quad{}=\ds -\sum\limits_{k\neq j}\beta_{jk}\bst_{ki}+(v_j-1)\bst_{ji}.
\end{array}
$$
Lemma is proved.
\end{dok}

The rotation coefficients of the \DE\ metric satisfy the following condition:
$\sum_k\p_k \beta_{ij}=0$. Our next goal is to prove the following
discrete analog of this property.
\begin{lem}\label{lb}
Let $\beta_{ij}$ be a solution of equations \lref{so1}{so3}, which satisfies 
(\ref{b}). Then it also satisfies the following monodromy property:
\begin{equation}\label{100}
\wT\beta_{ij}(u)=\beta_{ij}(u),\qquad \wT=\prod_{k=1}^n T_k.
\end{equation}
\end{lem}

\begin{dok}
First of all, we prove by induction that if functions $\Psi_i(u)$ 
satisfy (\ref{ll1}) then for any set~$I$ of pairwise distinct indices 
$I=\{i_1,\dots,i_s\}$ the following equation holds:
\begin{equation}\label{ll1a}
(T_I-1)\Psi_j=\sum_{i\in I}(T_I\beta_{ji})\Psi_i,\quad j\notin I,
\qquad T_I=T_{i_1}\ldots T_{i_s} .
\end{equation}
Indeed, if (\ref{ll1a}) is established for any set~$I$ of cardinality~$s$,
then for any $k\ne j$, $k\notin I$, we have
$$
\begin{array}{rcl}
(T_kT_I-1)\Psi_j&=&\ds T_I\D_k\Psi_j+(T_I-1)\Psi_j=T_I\bigl[
  (T_k\beta_{jk})\Psi_k \bigr]+\sum_{i\in I}(T_I\beta_{ji})\Psi_i={}\\[5mm]
{}&=&\ds (T_IT_k\beta_{jk})\Psi_k+(T_IT_k\beta_{jk})\bigl[
  (T_I-1)\Psi_k \bigr]+\sum_{i\in I}(T_I\beta_{ji})\Psi_i={}\\[5mm]
{}&=&\ds (T_IT_k\beta_{jk})\Psi_k+\sum_{i\in I} \bigl[
  T_I (T_k\beta_{jk}\beta_{ji}+\beta_{ji}) \bigr]\Psi_j={}\\[5mm]
{}&=&\ds (T_IT_k\beta_{jk})\Psi_k+\sum_{i\in I}(T_kT_I\beta_{ji})\Psi_i .
\end{array}
$$
The last equality proves the step of induction and completes the proof of
equation~(\ref{ll1a}).
Note, that the compatibility conditions for equations (\ref{ll1}) and
(\ref{ll1a}) lead to the following formula for the ``long'' difference
derivatives $\D_I=(T_I-1)$  of the discrete rotation coefficients:
$$
\D_I\beta_{jk}=\sum_{s\in I}(T_I\beta_{js})\beta_{sk},\qquad
j\neq k,\quad j,k\notin I .
$$
Let us also note, that as, due to (\ref{ss1}), the functions $\bst_{ij}$
for any index $k\neq i,j$ satisfy the same equations as $\Psi_i$,
we simultaneously prove the following equality:
\begin{equation}
\D_I\bst_{jk}=\sum_{s\in I}(T_I\beta_{js})\bst_{sk},\qquad j,k\notin I.
\label{ll1c}
\end{equation}
Equation (\ref{ll1a}) implies that
$$ 
\left( T^{(i)}-1 \right)\Psi_i=\sum_{j\neq i}
\left( T^{(i)}\beta_{ij} \right)\Psi_j,\qquad
T^{(i)}=\prod_{j\neq i}T_{j} .
$$ 
Since $(\wT-1)=T_i(T^{(i)}-1)+\D_i,$ applying (\ref{ll2}) we gain
the equality
$$
(\wT-1)\Psi_i=(\mu+B_i(u))\Psi_i+
\sum_{j=1}^n ((\wT-1)\beta_{ij})\Psi_j(u,\mu),
$$
where $B_i(u)$ is some function whose explicit form is irrelevant now.
Since the vectors~$\Psi_j$ are linearly independent, it now suffices 
to show that the vectors~$\Psi_i$ and $\wT\Psi_i$ are parallel.

Let us fix an arbitrary point $u_0$ and consider the solution
$\Psi_i(u)=\Psi_i(u; u_0)$ of equations \sref{ll1}{ll2} with the
following initial data at the point $u_0$:
\begin{equation}\label{m11}
\sca{\Psi_j(u_0),\Psi_k(u_0)}=\bst_{jk}(u_0),\qquad j,k=1,\ldots,n.
\end{equation}
These relations define the vectors $\Psi_i(u_0)$ uniquely up
to an orthogonal transformation of the whole space.

Let us prove by induction that relation (\ref{m11}) is satisfied at
the point $u_I=T_{i_1}\ldots T_{i_s} u_0$ for any set of pairwise distinct
indices $I=\{i_1,\ldots,i_s\}$ not containing $j$, \ie\
\begin{equation}\label{m11a}
\sca{\Psi_j(u_I),\Psi_k(u_I)}=\bst_{jk}(u_I),\qquad j\notin I.
\end{equation}
Suppose that $i\neq j,k$. Then
$$
\begin{array}{rcl}
\D_i\sca{\Psi_j,\ \Psi_k}&=&\sca{\D_{i}\Psi_j,\Psi_k} +
\sca{\Psi_j,\D_{i}\Psi_k}+\sca{\D_{i}\Psi_j,\D_{i}\Psi_k}={}\\[3mm]
{}&=&(T_i\beta_{ji})\sca{\Psi_i,\Psi_k}+(T_i\beta_{ki})\sca{\Psi_j,\Psi_i}+
(T_i\beta_{ji})(T_i\beta_{ki})\sca{\Psi_i,\Psi_i}.
\end{array}
$$
If the induction hypothesis is true for the point $u_I$ and $i\notin I$,
then we obtain the following formula for the scalar product at
the point $T_i u_I$:
\begin{equation}\label{m12}
\begin{array}{rcl}
\D_i\sca{\Psi_j(u_I),\Psi_k(u_I)}&=&(T_i\beta_{ji})\bst_{ik}+
  (T_i\beta_{k i})\bst_{ij}+(T_i\beta_{ki})(T_i\beta_{ji})\bst_{ii}={}\\[3mm]
{}&=&(T_i\beta_{k i})\bst_{ij}=\D_i\bst_{jk}
\end{array}
\end{equation}
(all the functions in the right hand side are evaluated at the point~$u_I$).
Analogously, using the induction hypothesis, we obtain
\begin{equation}\label{m45}
\begin{array}{c}
\D_i \sca{\Psi_j,\Psi_i}(u_I)=(T_i\beta_{ji})\bst_{ii}+
(\mu+v_i)\bst_{ij}-\sum_{k\neq i}\beta_{ik}\bst_{kj}+{}\\[5mm]
\ds\quad{}+(T_i\beta_{ji})\Bigl( (\mu+v_i)\bst_{ii}-
\sum_{k\neq i}\beta_{ik}\bst_{ki} \Bigr)=
(v_i-1)\bst_{ij}-\sum_{k\neq i}\beta_{ik}\bst_{kj}.
\end{array}
\end{equation}
(again, all the functions in the RHS are evaluated at the point~$u_I$).
Comparing the RHS of the last equality to the RHS of (\ref{ss3}), we obtain:
\begin{equation}\label{m46}
\D_i \sca{\Psi_j,\Psi_i}(u_I)=\D_i\bst_{ji}(u_I),\qquad
j\neq i,\quad j\notin I.
\end{equation}
Equations (\ref{m12}) and (\ref{m46}) imply (\ref{m11a}) at the 
point~$T_iu_I$.

Now we are ready to prove that the vectors $\Psi_i(u_0)$ and 
$\wT\Psi_i(u_0)$ are parallel. First, we show that the vectors 
$\Psi_m(u_I)$ and $T_mT_J\Psi_k(u_I)$ with non-intersecting
sets of indices $I,J\subset\{1,\dots,n\}$ are orthogonal provided
$m,k\notin I$, $m\notin J$ ($m\ne k$). According to (\ref{ll1a}), we have
\begin{equation}\label{92}
\sca{\Psi_m(u_I),T_mT_J\Psi_k(u_I)}=\sca{\Psi_m,\Psi_k}+\sum_{j\in J}
(T_mT_J\beta_{kj})\sca{\Psi_m,\Psi_j}+(T_mT_J\beta_{km})\sca{\Psi_m,\Psi_m}
\end{equation}
(all the functions in the RHS are evaluated at the point $u_I$).
We can apply formula (\ref{m11a}) to the scalar products in the RHS, so
$$
\begin{array}{rcl}
\sca{\Psi_m(u_I),T_mT_J\Psi_k(u_I)}&=&\ds\bst_{mk}+
\sum_{j\in J} (T_mT_J\beta_{kj})\bst_{jm}+(T_mT_J\beta_{km})\bst_{mm}={}\\[5mm]
{}&=&\ds\bst_{mm} T_m\Bigl( \D_J\beta_{km}-
\sum_{j\in J} (T_J\beta_{kj})\beta_{jm} \Bigr)=0 .
\end{array}
$$
This fact implies in particular, that the vectors
\begin{equation}\label{bas}
\Psi_1(u_0),\ T_1\Psi_2(u_0),\ T_1T_2\Psi_3(u_0),\ \ldots,\
T_1T_2\ldots T_{n-1}\Psi_n(u_0)
\end{equation}
form an orthogonal basis. On the other hand, considering the sets of indices
$I=\{1,2,\dots,s\}$, $m=s+1$, $J=\{s+2,\dots,n\}$ and $k=1$ for $s$ running
from $1$ to $n-1$, we establish that~$\wT \Psi_1(u_0)$ is orthogonal to all
the vectors of this basis, but for the first one.
This implies that the vectors $\Psi_1(u_0)$ and $\wT\Psi_1(u_0)$ are parallel. 
In the same manner we can show that the vectors $\Psi_i(u_0)$ and 
$\wT\Psi_i(u_0)$ are parallel for $i\ne 1$. As it was noticed above, 
it implies that $\beta_{ij}(u_0)=\wT \beta_{ij}(u_0)$. As the choice of 
the initial point $u_0$ was arbitrary, it completes the proof of the lemma.
\end{dok}

The definition of $\bst_{ij}$ and (\ref{100}) imply that
\begin{cor}
There exists a constant $\eta^2$  such that the following equations hold:
\begin{equation}
\wT\bst_{ij}=\eta^2 \bst_{ij}. \label{mon3}
\end{equation}
\end{cor}

Now we are ready to complete the proof of Theorem 3.2.

\begin{lem}\label{lb2}
Let functions $\beta_{ij}$ satisfy the conditions of Theorem 3.2 and let
$\eta^2$ be the corresponding Bloch multiplier, defined by (\ref{mon3}).
Then for $\mu=\eta-1$ there exists a solution $\Psi_i(u)$ of equations 
(\ref{ll1}) and (\ref{ll2}), satisfying relations (\ref{m11}) identically 
for $u$, \ie\
\begin{equation}\label{m11b}
\sca{\Psi_j(u),\Psi_k(u)}=\bst_{jk}(u),\qquad j,k=1,\dots,n.
\end{equation}
\end{lem}
\begin{dok}
Consider the solution $\Psi_i(u;u_0)$ for some point $u_0$. By definition 
of this solution, it satisfies the relations (\ref{m11b}) at the point~$u_0$. 
Let us show that if these relations are satisfied at the point~$u$ they are 
also satisfied at the point~$T_iu$. Suppose, for instance, that $i=1$. 
While proving lemma \ref{lb} we have shown that at the point $(T_1u)$  
equations (\ref{m11}) are satisfied for all the pairs of indices except 
for $(j=1, k=1)$. Therefore, we only have to show that
$$
|T_1\Psi_1|^2=\sca{T_1\Psi_1,T_1\Psi_1}=T_1\bst_{11}.
$$
The fact that vectors (\ref{bas}) form an orthogonal basis implies that
\begin{equation}\label{sk1}
|T_1\Psi_1|^2=\frac{\sca{T_1\Psi_1,\Psi_1}^2}{|\Psi_1|^2}+
\frac{\sca{T_1\Psi_1,T_1\Psi_2}^2}{|T_1\Psi_2|^2}+\ldots +
\frac{\sca{T_1\Psi_1,T_1\ldots T_{n-1}\Psi_n}^2}%
{|T_1\ldots T_{n-1}\Psi_n|^2}.
\end{equation}
Analogously to the derivation of equation (\ref{m45}), we obtain:
\begin{equation}\label{sk2}
\sca{T_1\Psi_1,\Psi_1}=(v_1+\mu+1)\bst_{11}-
\sum_{j\neq 1}\beta_{1i}\bst_{i1}=(\mu+1)\bst_{11}.
\end{equation}
From (\ref{m11a}) it follows that
\begin{equation}\label{sk3}
\sca{T_1\Psi_1,T_1\Psi_2}=T_1\bst_{12}.
\end{equation}
Besides, repeating the proof of (\ref{92}), we obtain
\begin{equation}\label{sk4}
\sca{T_1\Psi_1,T_1\ldots T_i\Psi_{i+1}}=T_1 \Bigl(
\sum_{p=2}^i (T_2\ldots T_i\beta_{i+1,p}) \bst_{p1}+\bst_{i+1,1} \Bigr)=
T_1\ldots T_i \bst_{1,i+1}
\end{equation}
(to get the last formula in (\ref{sk4}) we use (\ref{ll1c})).
Plugging expressions \lref{sk2}{sk4} in (\ref{sk1}), we have
$$
|T_1\Psi_1|^2=\frac{(\mu+1)^2(\bst_{11})^2}{\bst_{11}}+
\frac{(T_1\bst_{12})^2}{T_1\bst_{22}}+\ldots+
\frac{(T_1\ldots T_{n-1}\bst_{1n})^2}{T_1\ldots T_{n-1}\bst_{nn}}.
$$
Now, applying equations (\ref{ss2}) and (\ref{ss1}), we transform
the last expression to the form
$$
\begin{array}{rcl}
|T_1\Psi_1|^2&=&(\mu+1)^2\bst_{11}-T_1[ (T_2\beta_{12}) \bst_{12}]-
  \ldots-T_1\ldots T_{n-1}[ (T_n\beta_{1n}) \bst_{1n} ]={}\\[3mm]
{}&=&(\mu+1)^2\bst_{11}-T_1\D_2\bst_{11}-\ldots-
  T_1\ldots T_{n-1}\D_n\bst_{11}={}\\[3mm]
{}&=&(\mu+1)^2\bst_{11}+T_1\bst_{11}-\wT\bst_{11}=T_1\bst_{11} .
\end{array}
$$
Therefore, we prove that $\Psi_i(u;u_0)$ satisfy relations (\ref{m11b})
in the positive octant with the origin at~$u_0$.

Note, that the solutions $\Psi_i(u;u'_0)$ and $\Psi_i(u;u''_0)$ coincide 
in the intersection of the corresponding octants up to a general orthogonal 
transformation. We kill this freedom by fixing the solution $\Psi(u,0)$ and
choosing the initial conditions for $\Psi_i(u;u_0)$ with any $u_0$ so that 
$\Psi_i(u,u_0)$ coincides with $\Psi_i(u,0)$ in the intersection of their
domains. Then the function $\Psi_i(u)=\Psi_i(u;u)$ is a well-defined
solution of system \sref{ll1}{ll2} over the whole space, satisfying 
conditions (\ref{m11b}) for all $u$. Lemma is proved.
\end{dok}

\begin{cor} The above-constructed vector-functions $\Psi_i(u)$
satisfy the relation
\begin{equation}\label{ort}
\sca{T_j\Psi_i(u),\Psi_j(u)}=0,\qquad i\ne j .
\end{equation}
\end{cor}

Let functions $\beta_{ij}(u)$ satisfy the conditions of the theorem.
We define functions $h_i(u)$ as a solution to the system
\begin{equation}\label{h11}
\D_i h_j(u)=\beta_{ij}(u) T_ih_i(u),\qquad i\neq j .
\end{equation}
These equations are compatible due to (\ref{so1}). A solution of (\ref{h11}) 
depends on $n$ arbitrary functions of one variable, which are the initial
data, \ie\ functions $h_i(0,\dots,0,u_i,0,\dots,0)$.

Let us consider vector-functions
\begin{equation}\label{X}
X_i(u)=T_ih_i(u)\Psi_i(u),
\end{equation}
where $\Psi_i(u)$ were defined in the preceding lemma.
Equation (\ref{ll1}) implies
\begin{equation}\label{pla}
\D_jX_i=\left( T_i\frac{\D_jh_i}{h_i} \right) X_i+
\left( T_j\frac{\D_ih_j}{h_j} \right) X_j=\D_iX_j.
\end{equation}
Therefore, there exists a vector-function $\vx(u)$ such that
$X_i(u)=\D_i \vx(u)$. Due to (\ref{pla}) the function $\vx(u)$ defines
the planar lattice which, according to (\ref{ort}), is a \DE\ lattice.

To complete the proof of Theorem 3.2 we only need to show that
the functions $\beta_{ij}(u)$ are gauge equivalent to the rotation
coefficients of this lattice under transformation (\ref{so5}).

Let $\wt h_i(u)$ be the Lam\'e coefficients which, according to the
results of the previous section, correspond to the constructed lattice.
Equation (\ref{l1}) implies that $X_i$ satisfy the same equation (\ref{pla}),
where coefficients depend on the $\wt h_i$'s instead of the $h_i$'s. 
Therefore,
$$
\frac{\D_jh_i}{h_i}=\frac{\D_j\wt h_i}{\wt h_i}.
$$
Hence, $h_i=f_i\wt h_i$, where $f_i=f_i(u_i)$ depends only on the
variable~$u_i$.

Plugging (\ref{X}) into (\ref{ll2}) and using (\ref{h11}), we obtain:
$$
T_i X_i=\frac{T_i^2h_i}{T_ih_i} \left( (\eta+v_i)X_i-
\sum_{j\neq i} \frac{\D_ih_j}{T_jh_j}X_j \right).
$$
Equation (\ref{l2}) implies the same equality, where $h_i$ are replaced
by $\wt h_i$ and $\eta$ by $1/2$. Therefore, comparing the coefficients
of~$X_j$ we see that $f_i=c^{u_i}$, where $c$ is a constant.
The comparison of the coefficients at $X_i$ defines this constant:
$c=(2\eta)^{-1}$. Theorem 3.2 is proved.
\end{dok}

In the next section we present an algebro-geometric construction of a 
wide class of the \DE\ lattices, which can be written explicitly in terms 
of the theta-functions of auxiliary Riemann surfaces.

\section{Algebro-geometric lattices}

Let $\Gamma_0$ be a smooth genus $g_0$ algebraic curve on which there is a 
meromorphic function $E(P),$ $P\in\Gamma_0$, with $n$ simple poles and $n$ 
simple zeros. Let points $P_1,\dots,P_n$ be poles, and $Q_1,\dots,Q_n$ 
be zeros of $E(P)$. Consider the Riemann surface $\Gamma$ of the function 
$\sqrt{E(P)}$. It is two-sheeted covering of $\Gamma_0$ with $2n$ branching
points at the poles and the zeros of $E(P)$. According to the 
Riemann\,--\,Hurvitz formula genus $g$ of $\Gamma$ equals $g=2g_0+n-1$. 
Let $\sigma\colon\Gamma\to\Gamma$ be the holomorphic involution of $\Gamma$ 
which permutes sheets of the covering. The points $P_i$ and $Q_j$ are 
fixed points of the involution.

The function $E(P)$ on $\Gamma_0$ takes each value $n$ times.
Let us fix a constant $c^2$ and consider the points $P_i^c\in\Gamma_0$,
$i=1,\dots,n,$ such that $E(P_i^c)=c^2$.
The function $\lambda=c^{-1}\sqrt{E(P)}$ is odd with respect to the
involution $\sigma$, has simple poles at $P_1,\dots,P_n$ and simple zeros
at $Q_1,\dots,Q_n$. Let $P_i^{\pm}$ be preimages on  $\Gamma$ of the point
$P_i^c$. Then $\lambda(P_i^{\pm})=\pm 1$ and $\sigma(P_i^+)=P_i^-$.

We choose $w_i^+=\lambda-1,$  as a local coordinate on $\Gamma$ near $P_i^+$
and $w_i^-=\lambda+1$ as a local coordinate near $P_i^-$.
Note that $\sigma(w_i^+)=-w_i^-$.

Let us fix an integer $l\geq 1$ and two sets of points in the general
position on $\Gamma$. One of them is a set of $l$ points 
$\handR=(R_1,\dots,R_l)$, and the other is a set of $g+l-1$ points 
$\handD=(\gamma_1,\dots,\gamma_{g+l-1})$.
A pair of the divisors  $\handR,$ $\handD$ is called admissible
(see \cite{kr}), if there is a meromorphic differential $d\Omega_0$ on
$\Gamma_0$ with the following properties:

$(a^0)$\ $d\Omega_0$ has $n+l$ simple poles at $Q_i$,
$i=1,\dots,n$ and at the points $\wh R_{\alpha}$, which are
the projections of $R_{\alpha}$ on $\Gamma_0$, $\alpha=1,\dots,l$;

$(b^0)$\ the differential $d\Omega_0$ has $g+l-1$ zeros at the
projections $\wh\gamma_s$ of the points $\gamma_s$, $s=1,\dots,g+l-1$.

The differential $d\Omega_0$ may be considered as an even,
with respect to $\sigma$, differential on  $\Gamma$, where it has:

$(a)$\ $n+2l$ simple poles at $Q_i$, $i=1,\dots,n$, and at the points
$R_{\alpha}$ and $\sigma(R_{\alpha})$, $\alpha=1,\dots,l$;

$(b)$\ $2(g+l-1)$ zeros at $\gamma_s$ and $\sigma(\gamma_s)$,
$s=1,\dots,g+l-1$, and simple zeros at the points $P_i$, $i=1,\dots,n$.

The Riemann\,--\,Roch theorem implies that for each pair of divisors
$\handD,\handR$ in the general position there exists a unique
function $\psi(u,Q|\handD,\handR)$,
$u=(u_1,\dots,u_n)\in\inte^n$, $Q\in\Gamma$, such that:

$(1)$ $\psi(u,Q|\handD,\handR)$ as a function of the variable $Q$ is
meromorphic on $\Gamma$. Outside the punctures $P_i^{\pm}$ it has at most
simple poles at the points of the divisor $\handD$
(if all of them are distinct);

$(2)$ in the neighborhood of $P_i^{\pm}$ the function
$\psi(u,Q|\handD,\handR)$ has the form
\begin{equation}\label{ba1}
\psi=(w^{\pm}_i)^{\mp u_i}\left(
\sum_{s=0}^{\infty} \xi_{s,\pm}^i(u) (w^{\pm}_i)^s \right) ;
\end{equation}

$(3)$ $\psi$ satisfies the normalization conditions
\begin{equation}\label{ba2}
\psi(u,R_{\alpha}|\handD,\handR)\equiv 1,\qquad \alpha=1,\dots,l .
\end{equation}
The function $\psi$ is a discrete analog of the \BA\ functions which are
the core of algebro-geometric integration theory of soliton equations.
Further on, we shall often omit indication of its explicit dependence on
the divisors $\handD,\handR$ and will simply denote it by $\psi(u,Q)$.  
The discrete \BA\ function can be expressed in terms of the Riemann 
theta-function in a way almost identical to the continuous case 
(see \cite{kr}).

The Riemann theta-function, associated with an algebraic curve~$\Gamma$ 
of genus~$g$ is an entire function of $g$ complex variables
$z=(z_1,\dots,z_g)$, and is defined by its Fourier expansion
$$
\theta(z_1,\dots,z_g)=\sum\nolimits_{m\in\inte^g}
e^{2\pi i(m,z)+\pi i (Bm,m)},
$$
where $B=B_{ij}$ is a matrix of $b$-periods, $B_{ij}=\oint_{b_i}\omega_j$,
of normalized holomorphic differentials~$\omega_j(P)$ on~$\Gamma$:
$\oint_{a_j} \omega_i=\delta_{ij}$.
Here $a_i, b_i$ is a basis of cycles on $\Gamma$ with the canonical matrix
of intersections: $a_i\circ a_j=b_i\circ b_j=0$, $a_i\circ b_j=\delta_{ij}$.

The theta-function has the following monodromy properties with respect to
the lattice $\handB$, spanned by the basis vectors $e_i\in \comp^g$ and
the vectors $B_j\in \comp^g$ with coordinates $B_{ij}$:
$$
\theta(z+l)=\theta(z),\qquad
\theta(z+Bl)=\exp[-i\pi (Bl,l)-2i\pi (l,z)]\,\theta(z)
$$
where $l$ is an integer vector,  $l\in\inte^g$.
The complex torus $J(\Gamma)=\comp^g/\handB$ is called the Jacobian variety 
of the algebraic curve~$\Gamma$. The vector $A(Q)$ with coordinates
$$
A_k(Q)=\int_{q_0}^Q \omega_k
$$
defines the so-called Abel transform: $\Gamma\mapsto J(\Gamma)$.

According to the Riemann\,--\,Roch theorem, for each divisors
$\handD=\gamma_1+\ldots+\gamma_{g+l-1}$ and $\handR=R_1+\ldots+R_l$
in the general position there exists a unique meromorphic function
$r_{\alpha}(Q)$ such that $\handD$ is its poles' divisor and
$r_{\alpha}(R_{\beta})=\delta_{\alpha\beta}$. 
It can be written in the form
(see details in \cite{kr}):
$$
r_{\alpha}(Q)=\frac{f_{\alpha}(Q)}{f_{\alpha}(R_{\alpha})},\qquad 
f_{\alpha}(Q)=\theta(A(Q)+Z_{\alpha}) 
\frac{\prod_{\beta\neq \alpha} \theta(A(Q)+F_{\beta})}{%
\prod_{m=1}^l\theta (A(Q)+S_m)}, 
$$
where
$$
F_{\beta}={}-\handK-A(R_{\beta})-\sum_{s=1}^{g-1} A(\gamma_s),\qquad
S_m={}-\handK-A(\gamma_{g-1+m})-\sum_{s=1}^{g-1} A(\gamma_s),
$$
$$
Z_{\alpha}=Z_0-A(R_{\alpha}), \quad Z_0={}-\handK-\sum_{s=1}^{g+l-1}
A(\gamma_s)+\sum_{\alpha=1}^l A(R_{\alpha}),
$$
where $\handK$ is the vector of Riemann constants.

Let $d\Omega_j$ be a unique meromorphic differential on $\Gamma$, which is
holomorphic outside $P_j^{\pm}$, has simple poles at these punctures
with residues $\mp 1$, and is normalized by conditions
$$
\oint_{a_k}d\Omega_j=0 .
$$
It defines a vector $V^{(j)}$ with the coordinates
$$
V^{(j)}_k=\frac{1}{2\pi i} \oint_{b_k} d\Omega_j. 
$$
The \BA\ function $\psi(u,Q|\handD,\handR))$ has the form:
$$
\psi=\sum_{\alpha=1}^l r_{\alpha}(Q)
\frac{\theta(A(Q)+\sum_{i=1}^n (u^i V^{(i)})+Z_{\alpha})\;\theta(Z_0)}{%
\theta(A(Q)+Z_{\alpha})\;\theta(\sum_{i=1}^n (u^i V^{(i)})+Z_0)}
\exp{\left(\sum_{i=1}^n u^i\int_{R_{\alpha}}^Qd\Omega_i\right)} 
$$

\begin{teo}\label{tsch}
The \BA\ function $\psi(u,Q)$ satisfies the equation
\begin{equation}\label{sh}
\D_i\D_j\psi(u,Q)=a^i_{ij}(u)\D_i\psi(u,Q)+a^j_{ij}(u)\D_j\psi(u,Q),
\end{equation}
where
$$
a^i_{ij}=\frac{\D_jT_i\xi_{0,+}^i}{T_i\xi_{0,+}^i},\qquad
a^j_{ij}=\frac{\D_iT_j\xi_{0,+}^j}{T_j\xi_{0,+}^j}.
$$
and $\xi_{0,+}^s$ is the leading coefficient of expansion (\ref{ba1})
of $\psi(u,Q)$ near the puncture $P_s^+$.

\end{teo}
\begin{dok}
From the definition of the coefficients $a_{ij}^i$ it follows that the 
difference of the right and left hand sides of (\ref{sh}) satisfies the
first two defining properties of the Baker-Akhiezer function.
At the same time this difference equals zero at the points $R_{\alpha}$.
The uniqueness of the \BA\ function implies then that this difference
equals zero identically.
\end{dok}

Let $\eta^2_k=\Res_{Q_k}d\Omega_0$. 
Then we define a lattice $\vx(u)=(x_1(u),\dots,x_n(u))$ by the formula
\begin{equation}\label{lat}
x^k(u)=\eta_k\psi(u,Q_k) .
\end{equation}
Edges of this lattice are vectors $X_i(u)$, with coordinates
$(X_i)^k(u)=\eta_k\D_i\psi(u,Q_k)$. Let us define also vectors
$X_i^-(u)$ by the formula $(X_i^-)^k=-T_i^-(X_i)^k=\eta_k\D_i^-\psi(u,Q_k)$.
Evaluations of (\ref{sh}), at the points~$Q_k$, show that the above-defined
lattice is planar.

In a generic case the vector function ${\bf x}(u)$ is complex. Let us
specify algebro-geometric data that are sufficient for getting real vectors.

Suppose that on $\Gamma$ there exists an antiholomorphic involution~$\tau$,
such that it commutes with $\sigma$ and 
\begin{equation}\label{re}
\tau(\handD)= \handD,\quad \tau(\handR)=\handR,\quad
\tau(Q_i)=Q_i,\quad \tau(P_j^+)=P_j^+ .
\end{equation}
From the definition of $d\Omega_0$ it follows that
$\tau^*d\Omega_0=\ov{d\Omega_0}$. Therefore, the residue $\eta_k^2$
of that differential at ~$Q_k$ is a real number.
Suppose in addition, that $\eta_k$ is real (or $\eta^2_k$ is positive).
From (\ref{re}) it follows that defining analytical properties
of $\psi(u,Q)$, coincide with the analytical properties of
$\ov{\psi(u,\tau(Q))}$. Uniqueness of the \BA\ function implies then
that these functions coincide, \ie\ $\ov{\psi(u,\tau(Q))}=\psi(u,Q)$. 
Hence, $\vx(u)=\eta_k\bigl(\psi(Q_1,u),\dots,\psi(Q_n,u)\bigr)=\ov\vx(u)$,
and therefore, the lattice constructed  is a lattice in the real
Euclidian space.

\begin{lem}
Let $X_i^{\pm}(u)$ be vectors, which are defined by the \BA\ function. Then
$$
\sca{X_{i}(u),X_{j}^{-}(u)}=-\delta_{ij}\,(T_ih_i^{+}(u))\,(T_i^-h_i^{-}(u)),
$$
where $h_i^{\pm}(u)=\eps_i\xi_{0,\pm}^{i}(u)$, and $\eps_i^2$ is
the leading term of an expansion of $d\Omega_0$ in terms of local
coordinate $w_i^+$ in the neighborhood of the puncture $P_i^+$.
\end{lem}
\begin{dok}
Let us consider the differential
$$
d\Omega_{ij}=(\lambda(Q)-1)\, \Delta_i\psi(u,Q)\,
\Delta^-_j\psi(u,\sigma(Q))\, d\Omega_0 .
$$
If $i\ne j$ then that differential has poles at~$Q_k$, $k=1,\dots,n$ only.
Indeed, poles of $\lambda(Q)-1$ at~$P_k$, $k=1,\dots,n$, cancel with zeros
of $d\Omega_0$. At the points $\gamma_s$ and $\sigma(\gamma_s)$ poles of
the product $\Delta_i\psi \Delta^{-}_j\psi^{\sigma}$, cancel with zeros of
$d\Omega_0$. At the points $R_{\alpha}$ and $\sigma(R_{\alpha})$ the poles
of~$d\Omega_0$ cancel with zeros of the product. At the points $P_k^{\pm}$, 
$k\ne i,j$, the pole of one of the functions $\Delta_i\psi(u,Q)$,
$\Delta^{-}_j\psi(u,\sigma(Q))$ cancel with the zero of the other one.
The same is true for the points $P_i^-$ and $P_j^-$. The product
$\Delta_i\psi \Delta^{-}_j\psi^{\sigma}$ has poles at $P_i^{+}$
and~$P_j^{+}$, but at these points the function $\lambda(Q)-1$ has zeros.

The sum of residues of a meromorphic differential on a compact Riemann 
surface is equal to zero. Therefore,
$$
\sum_{k=1}^{n} \Res_{Q_k} d\Omega_{ij}=
\sum_{k=1}^{n} \Delta_i \vx^k(u)\cdot \Delta^{-}_j \vx^k(u) =
\sca{\Delta_i\vx(u), \Delta^{-}_j\vx(u)} = 0.
$$
Now let us consider the differential
$$
d\Omega_{ii}(u,Q)=
(\lambda(Q)-1)\,\D_i\psi(u,Q)\,\D_i^-\psi(u,Q^\sigma)\,d\Omega_0.
$$
It has additional pole at $P_i^+$ with the residue
$$
\Res_{P_i^+}d\Omega_{ii}=\eps_i^2\, (T_i\xi_{0,+}^{i})\,
(T_i^-\xi_{0,-}^{i}),
$$
which is equal to the sum of the residues at the punctures $Q_k$.
\end{dok}

Summarizing, we conclude that the lattice defined by (\ref{lat}) is
the \DE\ lattice. In order to show a complete correspondence with the
previous sections, let us find some other scalar products.
\begin{lem}
For scalar products of the vectors $X_i$ formulae \sref{m1}{m3}
$$
\sca{X_i,X_i} = 2(T_i h_i^{+})\, h_i^{-},\qquad
\sca{T_j X_i,X^{-}_i}=-(T_i T_j h_i^{+})\, (T^{-}_i h_i^{-}),
\quad i\neq j,
$$
where $h_i^{\pm}=\eps_i\xi_{0,\pm}^{i}$, are valid.
\end{lem}
\begin{dok}
Let us consider the differential
$$
d\Omega_{ii}^{+}=
\Delta_i\psi(u,Q)\, \Delta_i\psi(u,\sigma(Q))\, d\Omega_0 .
$$
It is meromorphic on~$\Gamma$ and has only simple poles at $Q_1,\dots,Q_n$,
and at $P_i^{\pm}$. Therefore, the sum of its residues $Q_1,\dots,Q_n$,
which coincides with the left hand side of (\ref{m1}), is equal to the sum
of residues at~$P_i^{+}$ and~$P_i^{-}$, taken with the negative sign.
We have
$$
\Res_{P_i^{+}} d\Omega^{+}_{ii}=\Res_{P_i^{-}} d\Omega_{ii}^{+}=
 (T_i h_i^{+}) (- h_i^{-}) ,
$$
which implies (\ref{m1}).

The proof of (\ref{m3}) is almost identical. It is enough to
apply the same arguments to the differential
$$
d\Omega_{ij}^{(1)}=
T_j\Delta_i\psi(u,Q)\, \Delta_i\psi(u,\sigma(Q))\, d\Omega_0 .
$$
The lemma is proved.
\end{dok}

Now let us define functions $\beta_{ij}^{\pm}(u)$ and $\bst_{ij}(u)$
by the formulae
$$
\beta_{ij}^{+}=\frac{\D_ih_j^+}{T_ih_i^+},\qquad
\beta_{ij}^{-}=\frac{\D_i^-h_j^-}{T_i^-h_i^-},\qquad
\bst_{ij}=\frac{\D_ih_j^-}{T_ih_i}.
$$
\begin{lem}
The functions  $\beta_{ij}^{\pm}(u)$ and $\bst_{ij}(u)$ satisfy
the equalities:
$$
\beta_{ij}^{+}=-\beta_{ji}^-,\qquad \bst_{ij}=\bst_{ji}.
$$
\end{lem}
\begin{dok}
Above we have proved formulae \lref{m1}{m3} for the vectors~$X_i$ and
the functions~$h_i^{\pm}$. Therefore, the statement of the lemma follows
from the results of Section 2. Nevertheless, we would like to show that
it can be directly proved within the framework of the algebro-geometric
construction.

The differential
$$
d\Omega_{ij}^{(2)}=
\lambda(Q)\, \D_j^-\psi(u,Q)\, \D_i\psi(u,\sigma(Q))\, d\Omega_0
$$
has simple poles at $P_j^-$ and $P_i^-$ only. The vanishing of the sum
of its residues at these punctures implies the first equality of the Lemma. 
To prove the second equality it is enough to consider the residues of 
the differential
$$
d\Omega_{ij}^{(3)}=
\lambda(Q)\, \D_j\psi(u,Q)\, \D_i\psi(u,\sigma(Q))\, d\Omega_0 ,
$$
which has only poles at $P_j^+$ and $P_i^+$.
\end{dok}

At the end, let us show that the functions
$$
\Psi_i(u,Q)=\frac{1}{T_i\xi_{0,+}^i}\D_i\psi(u,Q)
$$
satisfy equations which are gauge equivalent to (\ref{ll1}) and (\ref{ll2}). 
Note, that the vector-function $\Psi_i$ is uniquely defined by the following
analytical properties:

$(1)$ $\Psi_i(u,Q)$ as a function of $Q$ is meromorphic on $\Gamma$
and for each $u$ the divisor of its poles outside $P_i^{\pm}$
is less or equal to $\handD$;

$(2^+)$ in the neighborhood of $P_j^{+}$, the function $\Psi_i(u,Q)$
has the form
$$
\Psi_i=(w^{+}_j)^{-u_j-1}\left(
\delta_{ij}+\sum_{s=1}^{\infty} \zeta_{s,+}^j(u) (w^{+}_j)^s \right) ;
$$

$(2^-)$ in the neighborhood of $P_j^{-}$, the function $\Psi_i(u,Q)$
has the form
$$
\Psi_i=(w^{-}_j)^{u_j}\left(
\sum_{s=0}^{\infty} \zeta_{s,-}^j(u) (w^{-}_j)^s \right) ;
$$

$(3)$ $\Psi_i$ satisfies the normalization condition
$$
\Psi_i(u,R_{\alpha})\equiv 0,\qquad \alpha=1,\dots,l .
$$

\begin{teo}
The functions $\Psi_i(u,Q)$ satisfy the equations:
$$
\begin{aligned}
\D_j\Psi_i(u,Q)&=(T_j\gamma_{ij}(u))\Psi_j(u,Q), \qquad i\ne j, \\
\D_i\Psi_i(u,Q)&=(\mu+v_i)\Psi_i(u,Q)-
\textstyle\sum_{j\ne i}\gamma_{ij}(u)\Psi_j(u,Q),
\end{aligned}
$$
where
$$
\gamma_{ij}(u)=\frac{\D_i\xi_{0,+}^j(u)}{T_i\xi_{0,+}^i(u)},\qquad
v_i=-\sum_{j\ne i}\gamma_{ij}T_i\gamma_{ji}, \qquad
\mu(Q)=\frac{1}{\lambda(Q)-1}\,.
$$
\end{teo}
\begin{dok}
The proof of the Theorem is standard. The difference of the left and
right hand sides of the first eqaulity satisfies the first two properties
which define $\psi(u,Q)$, and equals zero at the normalization points.
Therefore, it equals zero identically. In the same way, the difference
of the left and right hand sides of the second equation is proportional
to $T_i\psi(u,Q)$. The evaluation of this difference at the normalization
points shows that it is identically zero.
\end{dok}

The coefficients $\gamma_{ij}$, defined in the theorem are connected with
the functions~$\beta_{ij}^{+}$ by the gauge transformation
$$
\beta_{ij}^+(u)=\frac{\eps_j}{\eps_i}\gamma_{ij}(u).
$$
\bigskip


\end{document}